%% file: main.tex
\documentclass[prd,twocolumn,superscriptaddress,showpacs,longbibliography,showkeys]{revtex4-2}
\usepackage[english]{babel} 
\usepackage{booktabs} 
\usepackage{enumitem} 
\setlist[itemize]{noitemsep} 
\usepackage[normalem]{ulem}

\usepackage[dvipsnames,usenames]{xcolor}
\usepackage{braket}
\usepackage{amsfonts}
\usepackage{amsmath}
\usepackage{amssymb}
\usepackage{mathtools}
\usepackage{bbold}
\usepackage{float}
\usepackage{physics}
\usepackage{tikz}
\usetikzlibrary{shapes,arrows}
\usepackage{hyperref}
\hypersetup{colorlinks,linkcolor={blue},citecolor={green},urlcolor={blue}}
\usepackage{qcircuit}
\usepackage{ulem}
\usepackage{subcaption}
\definecolor{cof}{RGB}{219,144,71}
\definecolor{pur}{RGB}{186,146,162}
\definecolor{greeo}{RGB}{91,173,69}
\definecolor{greet}{RGB}{52,111,72}

\newtheorem{defn}{Definition}[section]
\newtheorem{prop}{Property}[section]

\begin{document}

\title{An explicit tensor notation for quantum computing}

\newcommand{\UNITN}{{Dipartimento di Fisica, University of Trento, via Sommarive 14, 38123, Povo, Trento, Italy}}
\newcommand{\TIFPA}{Trento Institute for Fundamental Physics and Applications, Via Sommarive 14, 38123, Povo, Trento, Italy}

\author{Valentina Amitrano}
\affiliation{\UNITN}
\affiliation{\TIFPA}
\author{Francesco Pederiva}
\affiliation{\UNITN}
\affiliation{\TIFPA}

\date{\today}

\begin{abstract}
	This paper introduces a formalism that aims to describe the intricacies of quantum computation by establishing a connection with the mathematical foundations of tensor theory and multilinear maps. The focus is on providing a comprehensive representation of quantum states for multiple qubits and the quantum gates that manipulate them. The proposed formalism could contribute to a more intuitive representation of qubit states, and to a clear visualisation of the entanglement property.
	The main advantages of this formalism are that it preserves the fundamental structure of the Hilbert space to which quantum states belong, and also reduces the computational cost associated with classical prediction of the effect of quantum gates on multi-qubit states. 
	A connection between the ability to generate entanglement and the quantum gate representation is also established.
\end{abstract}

\maketitle

\section{Introduction}
Quantum computing has become one of the areas of physics in which the scientific community is investing the most effort. Starting with Feynman's idea~\cite{Feynman1982} and later with the foundations laid by Deutsch~\cite{deutsch1985}, we have arrived at applications that will change the way we approach computational~\cite{moll2018quantum, wittek2014quantum} and computer security problems~\cite{gisin2002quantum} and the way we solve the dynamics of quantum physical systems~\cite{Lloyd1073, klco2022standard, di2023quantum, bauer2023quantum, georgescu2014quantum}.  One of the main advantages of quantum computing is the possibility to manage amounts of data that grow exponentially with the number of available information units (qubits). However, this exponential growth unavoidably leads to a significant increase in the difficulty of designing algorithms that efficiently implement a given unitary transformation, and also makes it impossible to classically predict the results that a quantum computation should reproduce.
This is mainly related to the exponential growth affecting the representation of quantum states and operations on them. In turn, this is reflected in an increasing difficulty in using the so called quantum gate decomposition, a procedure required to implement a given unitary action on a system of multiple qubits in terms of a pre-determined set of unitary transformation organized into a quantum circuit that can be executed by a digital quantum computer~\cite{nielsen2001quantum}.
A realm of methods to perform quantum gate decomposition has been proposed and studied in the literature~\cite{barenco1995elementary, li2013decomposition, shende2005synthesis,tucci1999rudimentary, mottonen2004quantum,krol2022efficient, vartiainen2004efficient, vidal2004universal, vatan2004optimal, shende2003minimal, low2019hamiltonian, welch2014efficient, sawaya2020resource}. However, the scaling of the depth of the circuit generally remains exponential, and exact quantum gate decomposition becomes prohibitively difficult as the number of qubits increases, unless approximations are accepted~\cite{dawson2005solovay} or particular unitaries are considered. 

This exponential growth, that is common to quantum mechanics, is further complicated by the physical and computational notation used in the literature. However, a change in the formalism that has always been used could be useful to look at the science of quantum computing from a different perspective.
In particular, two notations are commonly used in the literature to describe quantum computing theory~\cite{nielsen2001quantum, yanofsky2008quantum}, both of which establish a convention for defining qubit states, gates and quantum algorithms. The first is the \textit{Dirac} notation, which has its origins in the formal expression of quantum mechanics in terms of Hilbert spaces and uses the \textit{ket} symbol $\ket{\psi}$ to denote the quantum state of the qubit system and operators to denote quantum gates.
This notation is intuitive, and is particularly useful for analysing algorithms that operate on many qubits. 

The other standard formalism is based on the so-called \textit{computational basis}, where qubit states are represented by vectors and quantum gates by unitary matrices. This notation is very useful when we use quantum computation in physical applications, such as the description of many-body quantum systems. Indeed, it allows one to predict the results of a quantum simulation in a classical way, by performing simple matrix multiplications, after having, of course, fixed a basis for our problem and mapped it onto the computational one. The main disadvantage of this notation is the exponential growth of vector and matrix dimensions, often making classical computing resources insufficient to predict the evolution of a number of qubits just over ten. 

The aim of this work is to develop an explicit \textit{tensorial notation}. 
This notation describes the qubit states as tensors and uses the concept of multilinear maps to implement quantum gates.
The main advantages of this formalism are that: a) it preserves the internal structure of the Hilbert space of qubits and operations on it, and b) it reduces the computational effort, so that any quantum transformation can be classically simulated by computing only $2 \times 2$ matrix multiplications.

This paper is structured as follows. 
Sec.~\ref{sec:math_background} contains an outline of the mathematical concepts used to introduce a tensor formalism in the field of quantum computing. 
Sec.~\ref{sec:QC_background} summarises the basic concepts of quantum computing and the two common notations used to describe them.
Sec.~\ref{sec:tensor_notation} presents the explicit tensor formalism for describing the state of multi-qubit systems and the quantum operations acting on them, pointing out the possibility of developing a new interpretation of qubit entanglement and the potential computational speedup that this formalism might allow.

\section{Mathematical background}
\label{sec:math_background}
We summarise here the key mathematical concepts and definitions~\cite{schlesinger2011algebra} needed to fix the notation and to understand the following sections. 

We define a vector space $V$ on a field $F \in \{\mathbb{R}, \mathbb{C}\}$ the set of elements equipped with two internal operations: the addition $+:V \times V \longrightarrow V$ and the scalar multiplication $\cdot: F \times V \longrightarrow V$.
Given two vector spaces $V$ and $W$ on $F$, with dimensions $\dim(V) = n$ and $\dim(W) = m$, we define a \textit{linear map} the application $f:V \longrightarrow W$.
The existing isomorphism
\begin{equation}
	Hom(V,W) \cong M_{m\times n},
	\label{eq:isomorphism}
\end{equation}
between linear maps and the space of $m \times n$ matrices, means that we can represent any linear map by a matrix.
Given three vector spaces $V,\,W,\,K$ on a field $F$ we define a \textit{bilinear map} the application $\phi : V \times W \longrightarrow K$ and the same definition can be extended to \textit{multilinear maps} $\phi:V_1 \times \dots \times V_d \longrightarrow K$.

\subsection{Tensor product space and tensor rank decomposition}
A tensor product~\cite{landsberg2012tensors} is a map $ \otimes : V \times W \longrightarrow V \otimes W$ that maps a pair of elements of two vector spaces to another element of the tensor product space, namely
\begin{equation}
	\otimes : (v,w) \longmapsto v \otimes w \in V \otimes W,
\end{equation}
for all $v \in V$ and $w \in W$.
The space $V \otimes W$ is uniquely defined up to isomorphisms and its dimension is the product of the dimensions, namely $\dim(V \otimes W) = n \cdot m$.
The elements of the tensor product space of the form $v \otimes w \in V \otimes W$ are called \textit{simple tensors} or \textit{pure tensors} and they generate the tensor product space, i.e.
\begin{equation}
    V \otimes W = \text{span}\left\{v \otimes w,  \forall v \in V, \forall w \in W\right\}.
\end{equation}
The basis of the tensor product space is the tensor product of the elements of the two bases, namely, given $\mathcal{B}_V = \{e_i\}$ the basis of $V$ and $\mathcal{B}_W = \{f_i\}$ the basis of $W$, $\mathcal{B}_{V \otimes W} = \{e_i \otimes f_j\}$ is the natural basis of $V \otimes W$.
A tensor is, in general, a linear combination of simple tensors, i.e.
\begin{equation}
    A = \sum_i \alpha_i \left( v_i \otimes w_i \right) \in V \otimes W,
\end{equation}
for $v_i \in V$, $w_i \in W$ and $\alpha_i \in F$ and can be represented by a two-dimensional array, i.e. a $n \times m$ matrix.
To represent a tensor product
\begin{equation}
	A = \sum_i \alpha_i \left( v_i^1 \otimes \dots \otimes v_i^d \right) \in V_1 \otimes \dots \otimes V_d
	\label{eq:A_tensor}
\end{equation}
of more than two elements one can use hypermatrices, which are the coordinate representation of tensors. The number $d$ of spaces connected by the tensor product is called the \textit{order} of the tensor and will also be referred to as $n$ in the rest of the paper, as it will take on the meaning of the number of qubits.

The \textit{tensor rank decomposition} consists of writing a tensor as the shortest linear combination of simple tensors
\begin{equation}
    A = \sum_{i=1}^r ( a_i^1 \otimes ... \otimes a_i^d) \in V_1 \otimes \dots \otimes V_d,
\end{equation}
where $r = \rank(A)$ and $a_i^k \in V_k$. Note that we have defined the elements $a_i^k$ such that $\alpha_i = 1$ for all $i$. A simple tensor has $r = 1$ by definition. 

\subsection{Operations on tensors}
\label{sec:mathematical_operation}
A tensor $A$ is transformed into another tensor $B$ through a multilinear map~\cite{vannieuwenhoven2012new}, namely
\begin{multline}
    B = (M_1,\dots,M_d) \cdot A\\
    = \sum_i \alpha_i \left(M_1v_i^1\right) \otimes \dots \otimes \left(M_d v_i^d\right),
    \label{eq:operation_multilinear}
\end{multline}
where $M_k:V_k \longrightarrow V_k$ are linear maps, $d$ is the order of tensors $A$ and $B$, and we used the Eq.~\eqref{eq:A_tensor} to express the tensor $A$ on the natural basis of the tensor product space.
An action on a tensor can be represented equivalently as the application of a linear map obtained by the tensor product of linear maps, i.e.
 \begin{equation}
    B = M_1\otimes \dots \otimes M_d (A)
    := \sum_i \alpha_i \left( b_i^1 \otimes \dots \otimes b_i^d\right),
    \label{eq:operation_linear}
\end{equation}
where we have defined $b_i^k := M_k v_i^k$.
This is related to the following property. 

\begin{figure}
    \centering
    \includegraphics[width=8.5cm]{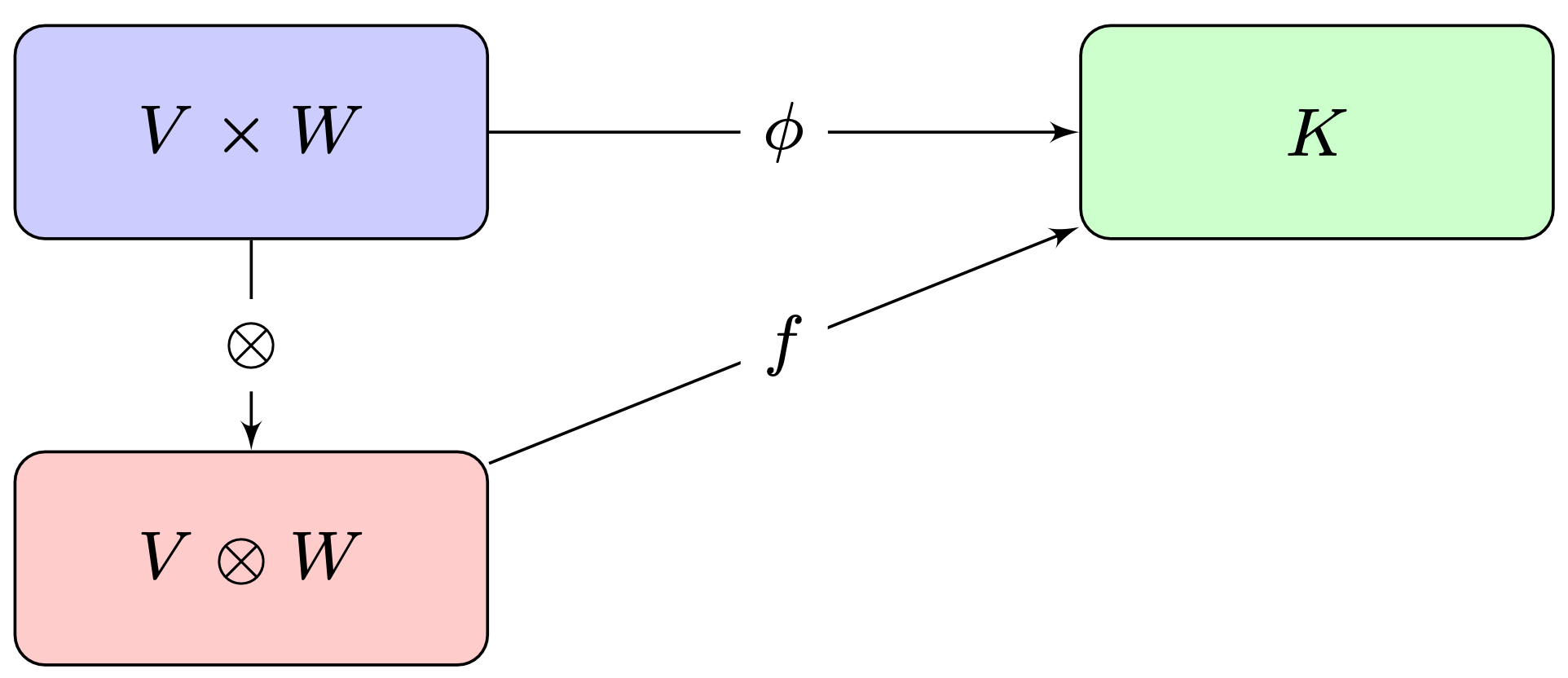}
    \caption{Sketch of the Property~\ref{prop:bilinear_tensor_linear} which allows to express a bilinear map $\phi$ as a linear map $f$.}
    \label{fig:bilinear_tensor_linear}
\end{figure}
\begin{prop}[Bilinear maps and tensor space]
\label{prop:bilinear_tensor_linear}
Given a bilinear map $\phi : V \times W \longrightarrow K$, there exists a unique linear map $f: V \otimes W \longrightarrow K$ such that
\begin{equation}
    \phi(v, w) = f \circ \otimes (v,w),
\end{equation}
where $v \in V$, $w \in W$ and $\phi(v,w) \in K$.
\end{prop}
The Property \ref{prop:bilinear_tensor_linear} implies that every bilinear map is a linear map after the correct embedding via the tensor product, and the result can be extended to multilinear maps $\phi : V_1  \times \dots \times V_d \longrightarrow K$ which can be represented by a linear map acting on the tensor product space, namely $f : V_1 \otimes \dots \otimes V_d \longrightarrow K$. The property \ref{prop:bilinear_tensor_linear} is represented by the scheme in the Figure \ref{fig:bilinear_tensor_linear}.

The following property will play a crucial role in the definition of quantum gates acting on qubit states expressed in tensorial notation.
\begin{prop}[Rank and operations on tensors]
\label{prop:rank_multilinear}
	The rank of a tensor does not increase under multilinear maps, namely
	\begin{equation}
   	 \rank((M_1,\dots,M_d) \cdot A) \leq \rank(A),
   	 \label{Prop:rank_multilinear}
	\end{equation}
	and remains the same if all matrices $ M_k $ have linearly independent columns, which is true, for example, if they are unitary matrices.
\end{prop}

\section{Quantum computing background and standard notations}
\label{sec:QC_background}
In the transition from classical to quantum computing, a new unit of information is defined. Instead of using the bit $b \in \mathbb{Z}_2 $ we use the qubit $\ket{q} \in \mathbb{C}^2$, thus changing the working space.
In Dirac notation we identify a single qubit state with a unit vector $\ket{\psi} \in \mathcal{H}_1$ in the Hilbert space $\mathcal{H}_1 = \mathbb{C}^2$, whose basis is $\mathcal{B}_1 = \{\ket{0}, \ket{1}\}$. The element satisfies the normalisation condition $\norm{\ket{\psi}}=1$, and the equivalence up to a global phase $\ket{\psi} \cong e^{i\varphi}\ket{\psi}$.

A state of $n$ qubits is an element of the tensor product Hilbert space $\mathcal{H}_{n} = (\mathbb{C}^2)^{\otimes n}$ of dimension $\dim(\mathcal{H}_{n}) = 2^n$ and its natural basis is given by the tensor product of the single qubit bases, i.e. 
\begin{equation}
\mathcal{B}_{n} =\left \{\ket{q_1}_1 \otimes \dots \otimes \ket{q_{n}}_n, \forall q_i \in \mathbb{Z}_2 \right\}.
\end{equation}
Each basis element can also be identified by $\ket{i}$, where $i$ is related to the decimal notation of the binary number $(q_n \dots q_{1})$ in reverse order, namely $i-1 = q_1 2^0 + q_2 2^1 + \dots + q_{n}2^{n-1}$. So we can also denote the $n$-qubit basis as 
\begin{equation}
	\mathcal{B}_n =\left \{\ket{i}, \forall i \in \{1,\dots,2^n\}\right\}.
	\label{eq:basis_decimal}
\end{equation}

Three main properties characterise quantum computing with respect to its classical counterpart. (1) Quantum superposition allows a single qubit to be in a linear combination of the basis elements, (2) the stochastic character of the measurement tells us that we can only access probabilistic information from a qubit state, and (3) entanglement allows different qubits to communicate in a non-local way and more information to be processed simultaneously.

According to the first property, the state of a single qubit can be in any linear combination of the basis elements of Hilbert space, namely $\ket{\psi} = \alpha \ket{0} + \beta \ket{1}$, where $\alpha, \beta \in \mathbb{C}$ and $\abs{\alpha}^2 + \abs{\beta}^2 = 1$.
This notation highlights the reason why a qubit is also defined as a \textit{two-level quantum system}.
Thus, from bit to qubit, there is an obvious increase in the information stored in the computational resource.

The probabilistic feature describes the fact that when we measure a qubit $\ket{\psi} \in \mathcal{H}_1$ we observe the state $\ket{0}$ with probability $\abs{\alpha}^2$ and the state $\ket{1}$ with probability $\abs{\beta}^2$. This means that we can only get probabilistic information about the state by measuring the same state $\ket{\psi}$ several times and sampling its probability distribution.
The post-measured state is a well-defined element of the basis $\mathcal{B}_1$ and the measurement procedure can therefore be viewed as a \textit{projection} along the basis. This physically causes the \textit{wave function collapse} which characterises the measurement as an \textit{irreversible} operation.

Finally the entanglement property allows distant qubits to be in a correlated state.
Using the multi-qubit basis $\mathcal{B}_{n}$ in Eq.~\eqref{eq:basis_decimal}, a multi-qubit state is a normalised linear combination
\begin{equation}
    \ket{\psi} = \sum_{i=1}^{2^n} \alpha_i \ket{i} \in  \mathcal{H}_n,
    \label{eq:psi_ket}
\end{equation}
where $ \alpha_i \in \mathbb{C}$ and $\sum_{i}\abs{\alpha_i}^2 = 1$.
According to the Schmidt decomposition theorem, there exists an orthonormal basis $\mathcal{B}_n^{Sch} = \{\ket{\phi_k}_1 \otimes \dots \otimes \ket{\phi_{k}}_n\}$ such that we can write the state as a linear combination of the minimum possible number of terms
\begin{equation}
    \ket{\psi} = \sum_{k=1}^s \beta_k \ket{\phi_k}_1 \otimes \dots \otimes \ket{\phi_k}_n \in \mathcal{H}_n,
    \label{eq:Schmidt_state}
\end{equation}
where $\beta_k$ coefficients satisfy the same properties of $\alpha_i$. In the above equation $s$ is called the \textit{Schmidt number} and we have $s=1$ if and only if $\ket{\psi}$ is a pure (separable) state and $s \geq 2$ implies that $\ket{\psi}$ is entangled.

\subsection{Qubit states as vectors}
\label{Sec:vector}
One of the most common ways to express the state of a qubit is based on the canonical basis of the Hilbert space $\mathcal{H}_1 = \mathbb{C}^2$, i.e.
\begin{equation}
    \mathcal{B}_1 = \left\{ \begin{pmatrix}
        1 \\ 0
    \end{pmatrix} , \begin{pmatrix}
        0 \\ 1
    \end{pmatrix} \right\} \subset \mathbb{C}^2.
    \label{eq:basis_computational_1}
\end{equation}
In this context a single qubit state is a two-dimensional normalised vector
\begin{equation}
    \ket{\psi} = \begin{pmatrix}
        \alpha \\ \beta
    \end{pmatrix} \in \mathbb{C}^2.
\end{equation}
Considering more than one qubit in the system, the standard notation consists in defining the \textit{computational basis} using exponentially large vectors based on the isomorphism
\begin{equation}
    (\mathbb{C}^2)^{\otimes n} \cong \mathbb{C}^{2^n}.
    \label{eq:isomorphism_space}
\end{equation}
This is achieved by means of the Kronecher product~\cite{henderson1983history} definition.
For example, the second basis element of an $n=2$ qubit system can be found by calculating
\begin{equation}
    \ket{01} = \begin{pmatrix} 1 \\ 0 \end{pmatrix} \boxtimes \begin{pmatrix} 0 \\ 1 \end{pmatrix} 
    = \begin{pmatrix} 0 \\ 1 \\ 0 \\ 0 \end{pmatrix}.
    \label{eq:basis_element_2}
\end{equation}
The generalisation to $n$ qubits is straightforward and the computational basis of the $n$ qubit Hilbert space $\mathcal{H}_n$ is
\begin{equation}
	\mathcal{B}_n = \left\{\delta_k, \forall k \in \{1,\dots,2^n\}\right\},
\end{equation}
where $\delta_k \in \mathbb{C}^{2^n}$ is a vector with 0 entries in rows $j \not = k$ and 1 entry in row $k$.
In summary, the Kronecker product between two tensors increases the dimensions while keeping the order of the tensor constant. For this reason, we can think of an $n$ qubit state as a tensor of order 1 and dimension $2^n$, i.e. a vector.

\subsection{Quantum gates as matrices}
\label{Sec:gate_matrix}
If we use vector notation to denote qubit states, we can act on them by matrix multiplication.
In particular, the $\mathcal{SU}(2)$ group completely describes the physics of any state $\ket{\psi} \in \mathcal{H}_1$. 
Its algebra $\mathfrak{su}(2)$, represented on the space $\mathbb{C}^2$ and using the canonical basis~\eqref{eq:basis_computational_1}, is generated by the Pauli matrices, represented by the following $2 \times 2$ hermitian, unitary, traceless, (-1)-determinant matrices
\begin{equation}
    \sigma_x = \begin{pmatrix}
        0 & 1 \\ 1 & 0
    \end{pmatrix},  \,
    \sigma_y = \begin{pmatrix}
        0 & -i \\ i & 0
    \end{pmatrix},  \,
    \sigma_z = \begin{pmatrix}
        1 & 0 \\ 0 & -1
    \end{pmatrix}.
\end{equation}
Each single qubit gate $G_i \in \mathcal{SU}(2)$ is a $2 \times 2$ unitary matrix which can be written as a linear combination of the elements in the Pauli group $\left\{ \mathbb{1}, \sigma_x, \sigma_y, \sigma_z \right\}$, and it is the matrix representation of a linear map $G_i : \mathbb{C}^2 \longrightarrow \mathbb{C}^2$. This follows from the property in Eq.~\eqref{eq:isomorphism}. In this sense, single-qubit gates belong to the space $\mathbb{C}^2 \otimes \mathbb{C}^2$. For the two-qubit case, quantum gates $G_{ij} \in \mathcal{SU}(4)$ are $4 \times 4$ unitary matrices, and a general operation on $n$ qubits can be expressed as a $2^n \times 2^n$ unitary.

Again, this notation comes from the definition of the Kronecker product. For example, consider two single-qubit gates $G_1, G_2 \in \mathcal{SU}(2)$, expressed as $2 \times 2$ matrices acting on two different qubits. The global two-qubit operation can be expressed as a $4 \times 4$ matrix, resulting from the Kronecker product of the two, namely
\begin{multline}
    G_1 \boxtimes G_2 = \begin{pmatrix}
    a_0 & a_1 \\ a_2 & a_3
    \end{pmatrix} \boxtimes \begin{pmatrix}
    b_0 & b_1 \\ b_2 & b_3
    \end{pmatrix} \\
    = \begin{pmatrix}
    a_0\begin{pmatrix}
    b_0 & b_1 \\ b_2 & b_3
    \end{pmatrix} & a_1\begin{pmatrix}
    b_0 & b_1 \\ b_2 & b_3
    \end{pmatrix} \\ a_2\begin{pmatrix}
    b_0 & b_1 \\ b_2 & b_3
    \end{pmatrix} & a_3\begin{pmatrix}
    b_0 & b_1 \\ b_2 & b_3
    \end{pmatrix}
    \end{pmatrix}.
\label{eq:2qubits_gate}
\end{multline}
So we can think of a two-qubit gate as an operation belonging to $(\mathbb{C}^2 \otimes \mathbb{C}^2) \boxtimes (\mathbb{C}^2 \otimes \mathbb{C}^2)$ and each two-qubit gate can be expressed as a sum of single Kronecker products as
\begin{equation}
    G_{12} = \sum_{i=1}^r \left( G_1^i \boxtimes G_2^i \right) \in \mathcal{SU}(4).
    \label{Eq:2qubit_gate_matrix}
\end{equation}
The notation can easily extended to an $n$-qubit gate as
\begin{equation}
    G = \sum_{i=1}^r \left( G_1^i \boxtimes \dots \boxtimes G_n^i \right) \in \mathcal{SU}(2^n).
    \label{Eq:nqubit_gate_matrix}
\end{equation}

The Kronecker notation is the most common one for classical prediction of the evolution of a qubit system.
However, this is only possible for a small number of qubits because the dimension of the vectors (qubit states) and the matrices (quantum gates) grows exponentially with the number $n$ of qubits involved. 
A disadvantage of this notation is precisely the exponential growth of the computational resources required to perform this matrix multiplication. 
Another disadvantage of the Kronecker-based notation is that it increases the non-trivial coefficients in the matrix (the $4 \times 4$ matrix in Eq.~\eqref{eq:2qubits_gate} has 16 non-trivial entries) and \textit{hides} the local property of quantum gates, as explained in the following section. 

\subsubsection{Local gate detection}
\label{Sec:local_gate_detection}
Given an $n$ qubit gate $G \in \mathcal{SU}(2^n)$, it is generally non-trivial to determine whether the gate is local or entangled by looking at the matrix notation, which has $4^n$ potentially non-trivial complex elements.
On the other hand, we know that a tensor is separable if and only if it is a rank-1 tensor.
At this point it is necessary to pay attention to the space in which the rank is computed and to what is meant by the separability of a tensor in relation to the separability of a gate.
For example, consider a local two-qubit gate $G_L = G_1 \boxtimes G_2$ like the one in Eq.~\eqref{eq:2qubits_gate}.
If we calculate the rank of its matrix representation we get $\rank(G_L) = 4$.
Thus, calculating the rank of a local quantum gate $G_L \in \mathcal{SU}(2^n)$ expressed in the common notation would always give $\rank(G_L) = 2^n$.
This is because we are not using the appropriate space on which to define the separability of a multi-qubit gate.
Returning to the definition of the two-qubit gate, it may be useful to define the tensor space of gates as $V_1 \otimes V_2$, where $V_1 = \mathbb{C}^2 \otimes \mathbb{C}^2$ is the space of gates acting on the first qubit and $V_2 = \mathbb{C}^2 \otimes \mathbb{C}^2$ on the second.
The locality character of a gate means that it acts independently on the qubits, so the separability of a two-qubit gate has to be defined in the space $V_1 \otimes V_2$. Elements of $\mathbb{C}^2 \otimes \mathbb{C}^2$ are $2 \times 2$ matrices, but elements of $V_1$ are vectors of dimension 4 due to the isomorphism $\mathbb{C}^2 \otimes \mathbb{C}^2 \cong \mathbb{C}^4$ (Eq.~\eqref{eq:isomorphism_space}).
A tensor product of two 4-dimensional vectors $G_1 \in V_1$ and $G_2 \in V_2$ is a tensor of order 2, i.e. a $4 \times 4$ matrix. 
Thus, in the $V_1 \otimes V_2$ space, the two-qubit local gate of Eq.~\eqref{eq:2qubits_gate}  is
\begin{multline}
    G_1 \otimes G_2 = \begin{pmatrix}
    a_0 \\ a_1 \\ a_2 \\ a_3
    \end{pmatrix} \begin{pmatrix}
    b_0 & b_1 & b_2 & b_3
    \end{pmatrix} \\
    = \begin{pmatrix}
    a_0 \begin{pmatrix} b_0 & b_1 & b_2 & b_3 \end{pmatrix}\\
    a_1 \begin{pmatrix} b_0 & b_1 & b_2 & b_3 \end{pmatrix}\\
    a_2 \begin{pmatrix} b_0 & b_1 & b_2 & b_3 \end{pmatrix}\\
    a_3 \begin{pmatrix} b_0 & b_1 & b_2 & b_3 \end{pmatrix}
    \end{pmatrix},
\end{multline}
whose columns are all linearly dependent, i.e. $\rank(G_L) = 1$, and therefore represents a separable gate.
In conclusion, using the space $(\mathbb{C}^2 \boxtimes \mathbb{C}^2) \otimes (\mathbb{C}^2 \boxtimes \mathbb{C}^2)$, we have established a one-to-one correspondence between separable gates and rank-1 matrix representations. The property can be extended to $n$ qubit gate locality using the rank definition of hypermatrices.

\section{The explicit tensor notation}
\label{sec:tensor_notation}
So far we have defined qubit states as vectors and quantum gates as matrices, both of which have exponentially increasing dimensions.
We now introduce a different way of describing these two elements, which could potentially give us a new perspective for unlocking the properties of multi-qubit systems and for reducing the amount of computational resources needed to classically predict the action of a quantum gate.

\subsection{Qubit states as tensors}
According to the Dirac notation in Eq.~\eqref{eq:psi_ket}, a multi-qubit state is an order-$n$ tensor which can be represented in coordinates as a hypermatrix.
However, the common notation uses the Kronecker product, which introduces the isomorphism in Eq.~\eqref{eq:isomorphism_space} and makes a multi-qubit state represented by an order-1 tensor, i.e. a vector, with exponentially increasing dimension.
This mixes the Hilbert spaces $\mathcal{H}_1$ associated with different qubits, often making the description non-intuitive.
If instead we keep the tensor product $\otimes$, instead of the Kronecker one $\boxtimes$, this increases the order by keeping each dimension constant, as opposed to increasing the dimension by keeping the order constant.

Let's start by considering a system with $n=1$ qubit. 
Since we are not doing a Kronecker product, the state $\ket{\psi} \in \mathbb{C}^2$ is the same in vector and tensor notation.
Using a pictorial representation, a general single-qubit state is a vertical line 
\begin{equation}
\begin{tabular}{c}
	\input{1_line}
	\end{tabular}
	\label{eq:1_line}
\end{equation}
where the upper vertex corresponds to $\ket{\psi} = \ket{0}$ and the lower vertex corresponds to $\ket{\psi} = \ket{1}$.
Going up to $n=2$, a two-qubit state belongs to the space $\mathcal{H}_2$, and in standard notation we represent it as $\ket{\psi} \in \mathbb{C}^4$. If we replace the Kronecker product by the tensor product, we get instead an order-2 tensor $\ket{\psi} \in \mathbb{C}^2 \otimes \mathbb{C}^2$, expressed as a linear combination of the basis elements in
\begin{equation}
    \mathcal{B}_2 = \left \{\begin{pmatrix}1 & 0\\0 & 0\end{pmatrix}, \begin{pmatrix}0 & 1\\0 & 0\end{pmatrix},
    \begin{pmatrix}0 & 0\\1 & 0\end{pmatrix}, \begin{pmatrix}0 & 0\\0 & 1\end{pmatrix}
    \right \}.
\end{equation}
The second basis element is, for example
\begin{equation}
        \ket{01} = \begin{pmatrix}1\\0 \end{pmatrix} \otimes \begin{pmatrix}0 \\ 1\end{pmatrix} = \begin{pmatrix}1\\0 \end{pmatrix}
        \begin{pmatrix}0 & 1\end{pmatrix} = \begin{pmatrix}0 & 1\\0 & 0\end{pmatrix},
\end{equation}
compared to the one in Eq.~\eqref{eq:basis_element_2}.
Using a pictorial representation, a general two-qubit state can be represented as the square
\begin{equation}
\begin{tabular}{c}
	\input{2_square}
\end{tabular}
\label{eq:2_cube}
\end{equation}
where we have added, to the vertical direction corresponding to the first qubit $\ket{q}_1$, an additional dimension (the horizontal one) representing the second qubit $\ket{q}_2$. The left vertex corresponds to $\ket{q}_2 = \ket{0}_2$ and the right one to $\ket{q}_2 = \ket{1}_2$.
Increasing the number of qubits to $n = 3$, a state $\ket{\psi} \in \mathbb{C}^2 \otimes \mathbb{C}^2 \otimes \mathbb{C}^2$ is represented on the computational basis as a vector $\ket{\psi} \in \mathbb{C}^8$ with eight elements. Replacing the Kronecker with the tensor product definition we get an order-3 tensor, i.e. a cube
\begin{equation}
\begin{tabular}{c}
	\input{3_cube}
\end{tabular}.
\label{eq:cube}
\end{equation}
The generalisation to an $n$ qubit system is straightforward and gives an order-$n$ tensor instead of a $2^n$ dimensional vector.

\subsubsection{Separable and entangled states}
The fact that this explicit tensorial notation avoids mixing Hilbert spaces of different qubits introduces two main advantages in the interpretation of qubit states. First of all, (1) it makes clear whether some qubits are in a basis state by fixing the portion of the hypermatrix corresponding to it.
Using the standard notation, this would require first computing the corresponding binary string for each non-zero entry of the vector representation.
For example, the separable three-qubit state $\ket{\psi} = (\alpha \ket{0}_1 + \beta \ket{1}_1) \otimes \ket{0}_2 \otimes \ket{0}_3$ is, in vector notation,
\begin{equation}
    \ket{\psi} 
    = \begin{pmatrix}\alpha & 0 & 0 & 0 & \beta & 0 & 0 & 0\end{pmatrix}^T,
\end{equation} 
where $\alpha$ and $\beta$ are in positions $i=1$ and $ j=5$ corresponding to the (reverse ordered) binary strings $[i-1] = (000)$ and $[j-1] = (100)$.
On the other hand, the corresponding order-3 tensor is
\begin{equation}
\begin{tabular}{ccc}
	$\ket{\psi}$ & = &
	\begin{tabular}{c}
	\begin{tikzpicture}[scale=0.8]
        \coordinate (A) at (1,-1,1);
        \coordinate (B) at (-1,-1,1);
        \coordinate (C) at (-1,1,1);
        \coordinate (D) at (1,1,1);
        \coordinate (E) at (1,1,-1);
        \coordinate (F) at (-1,1,-1);
        \coordinate (G) at (-1,-1,-1);
        \coordinate (H) at (1,-1,-1);
        \draw[fill=black] (C) circle (1pt) node[anchor=east] {$\alpha$};
        \draw[fill=black] (F) circle (1pt) node[anchor=east] {$0$};
        \draw[fill=black] (D) circle (1pt) node[anchor=west] {$0$};
        \draw[fill=black] (E) circle (1pt) node[anchor=west] {$0$};
        \draw[fill=black] (B) circle (1pt) node[anchor=east] {$\beta$};
        \draw[fill=black] (G) circle (1pt) node[anchor=east] {$0$};
        \draw[fill=black] (A) circle (1pt) node[anchor=west] {$0$};
        \draw[fill=black] (H) circle (1pt) node[anchor=west] {$0$};
        \draw[opacity=0.4] (A)--(B)--(C)--(D)--cycle;
        \draw[opacity=0.4] (E)--(F)--(G)--(H)--cycle;
        \draw[opacity=0.4] (A)--(H);
        \draw[opacity=0.4] (B)--(G);
        \draw[opacity=0.4] (D)--(E);
        \draw[opacity=0.4] (C)--(F);
        \end{tikzpicture}
	\end{tabular}
\end{tabular}
\end{equation}
which makes evident the three-qubit state.
The non-zero entries are in fact in the intersection between the front face ($\ket{q}_3 = \ket{0}_3$) and the left face ($\ket{q}_2 = \ket{0}_2$), while the vertical dimension ($\ket{q}_1$) represents a linear combination of the two states $\ket{0}_1$ (top face) and $\ket{1}_1$ (bottom face).

\begin{figure}
    \centering
    \includegraphics[width=8.5cm]{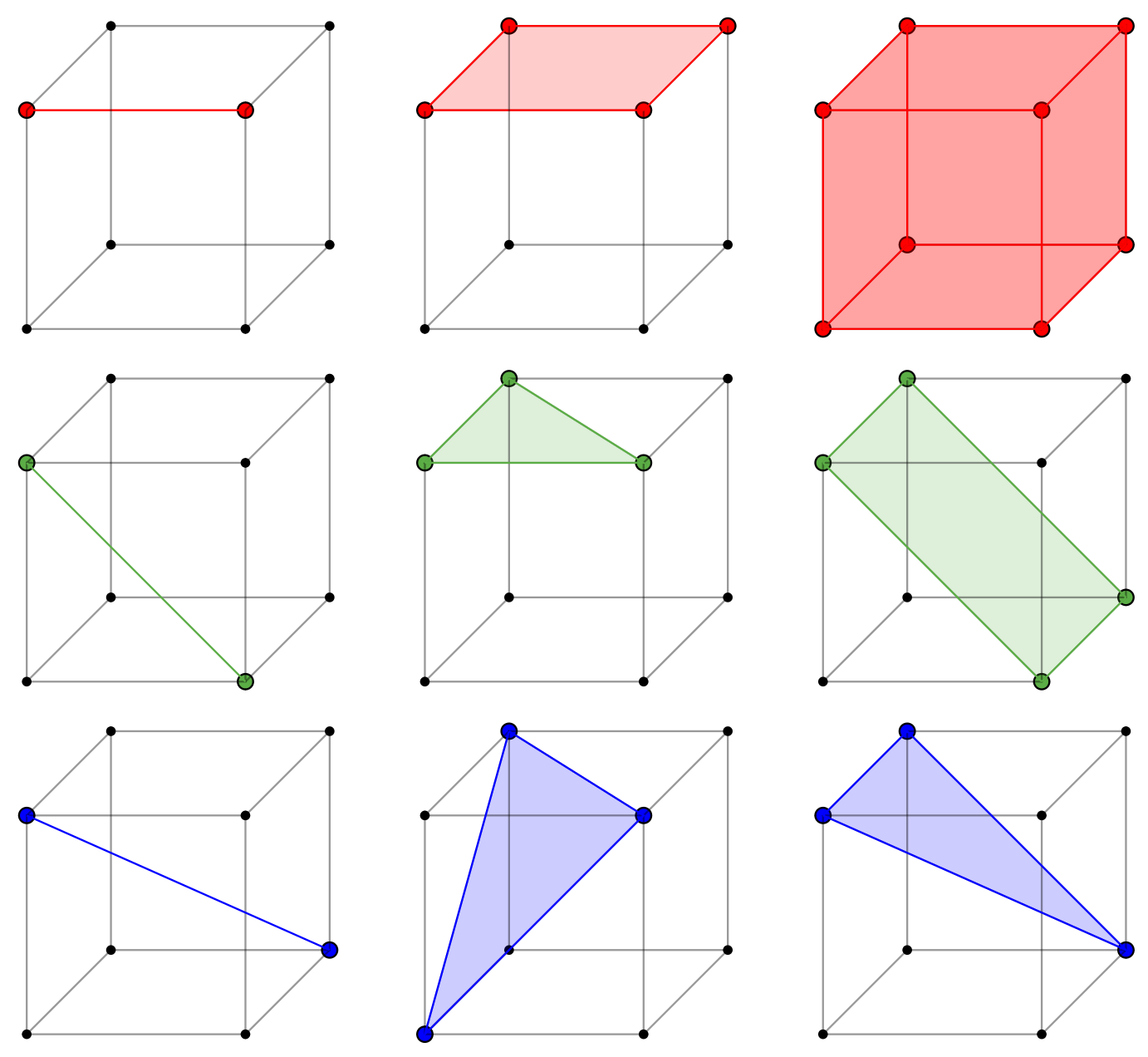}
    \caption{General shape representations of three-qubit states in tensor notation, showing separable states (first panel) $\ket{0}\otimes \ket{+} \otimes \ket{0}$, $\ket{0} \otimes \ket{+} \otimes \ket{+}$ and $\ket{+} \otimes \ket{+} \otimes \ket{+}$ from left to right, partially entangled states (second panel) $(\ket{00} + \ket{11})/\sqrt{2} \otimes \ket{0}$, $\ket{0}\otimes (\ket{00}+\ket{01}+ \ket{11})/\sqrt{3}$ and $(\ket{00}+\ket{11})/\sqrt{2} \otimes\ket{+}$ from left to right, and fully entangled states (bottom panel) $(\ket{000}+ \ket{111})/\sqrt{2}$, $(\ket{001}+\ket{010}+\ket{100})/\sqrt{3}$ and $(\ket{000}+ \ket{001}+ \ket{111})/\sqrt{3}$ from left to right. The figure highlights the relationship between the dimensionality of the entanglement and the connecting lines between non-trivial vertices.}
    \label{Fig:tensor_entanglement}
\end{figure}

Furthemore, (2) it makes clear whether a qubit state is separable or entangled by establishing the type of connection between portions of the hypermatrix.
The Figure~\ref{Fig:tensor_entanglement} shows the general cube representation of some separable, partially entangled and fully entangled 3-qubit states, where non-zero entries (i.e. $\alpha_i \not = 0$) are represented by bold coloured dots. 
The first panel in Fig.~\ref{Fig:tensor_entanglement} highlights the fact that separable states are represented by cubes with non-zero entries that can be connected by lines along the edges. On the other hand, the cubes representing the tensor of partially entangled states, such as those in the second panel of Fig.~\ref{Fig:tensor_entanglement}, have at least two non-zero vertices that must be connected along a diagonal. The surface to which the diagonal belongs defines the dimension along which the state is entangled. Each partially entangled state belongs to a plane that defines the separable qubit and its state. Finally, for tensors of fully entangled three-qubit states in the third panel of Fig.~\ref{Fig:tensor_entanglement}, we must cross each dimension along a diagonal to connect all non-zero vertices.
The number of non-zero entries obviously corresponds to the number of basis elements in the linear combination (as for the number of non-zero entries in vector notation), but the way in which we can connect them is related to the degree of entanglement of the state.
Using the standard notation one should first compute the Schmidt decomposition.

Many-qubit entanglement is a topic that has been widely studied in the literature~\cite{holweck2016three, holweck2017entanglement, gharahi2020fine} introducing different entanglement measures, such as~\textit{non-locality} \cite{de2021mermin, batle2011nonlocality}, \textit{geometric measure}~\cite{wei2003geometric}, \textit{tensor rank}, \textit{von Neumann entanglement entropy} and many more.
However, many-qubit entanglement still has no unique interpretation, since different entanglement measures generally fix different entanglement hierarchies. 
The explicitly tensorial representation presented in this work could provide an intuitive way to detect separable and entangled three-qubit states, and could introduce a new measure of entanglement for a general number of qubits $n$. This could be calculated by the number of off-edge connections needed to connect all non-zero vertices.
We believe that this could be potentially very interesting and have left its study to future work. 

\subsubsection{Measurement interpretation}
\label{sec:measurement}
The avoided Hilbert space mixture proposed by the tensor representation also provides a more intuitive interpretation of the measurement procedure of a multi-qubit state.
The hypercube representation of an $n$-qubit state highlights the probability of measuring a given basis element $\ket{i}$ as the absolute value of the complex number $\abs{\alpha_i}^2$ of the corresponding vertex. For example, the probability of obtaining the state $\ket{000}$ by measuring the tensor in Eq.~\eqref{eq:cube} is $\abs{\alpha}^2$. 
This is also true for vector notation, but to identify the position of a basis element along the vector, one should first calculate the corresponding bit string.
Moreover, measuring a single qubit from $n$ collapses the tensor of order $n$ to a tensor of order $n-1$, fixing the half of the tensor corresponding to the outcome measure. For example, measuring $\ket{q}_2 = \ket{0}_2$ in Eq.~\eqref{eq:cube} will fix the left face of the cube, while measuring $\ket{q}_3 = \ket{0}_3$ will fix the front face. The state of the unmeasured qubit $\ket{q}_1 = \alpha\ket{0}_1 + \beta \ket{1}_1$ is highlighted in the edge corresponding to the intersection of the two selected faces.
Once we have measured all the qubits in the system, we are left with a single basis element $\ket{i}$, and the probability of measuring it at the beginning of the measurement procedure was $\abs{\alpha_i}^2$.

\subsection{Quantum gates using multilinear map-based notation}
\label{sec:quantum_gate_tensor}
Once we have fixed the qubit state representation to tensors, we need a coherent way to act on them.
As mentioned in Sec.~\ref{sec:mathematical_operation}, operations on tensors can be implemented by multilinear maps (cf. Eq.~\eqref{eq:operation_multilinear}) or linear maps (cf. Eq.~\eqref{eq:operation_linear}) after the space embedding. The latter is the one used in the standard notation, which embeds the spaces $\mathcal{H}_1 = \mathbb{C}^2$ via the Kronecker product.
Similarly to the qubit states, the standard notation mixes the gates acting on different qubits increasing the number of non-trivial entries in the matrix (cf. Eq.~\eqref{eq:2qubits_gate}).
Moreover, this breaks the connection between the locality character and the rank of the matrix representation, as proved in Sec.~\ref{Sec:local_gate_detection}.
We now want to reformulate the notation used to represent gates by using the property~\ref{prop:bilinear_tensor_linear} in the opposite way, i.e. writing linear maps (quantum gates in standard notation) as multilinear maps (quantum gates in tensorial notation), thus avoiding the embedding.

\subsubsection{Local gates as multilinear maps}
Single qubit gates $G_i \in \mathcal{SU}(2)$, acting on the qubit $i$, are $2 \times 2$ unitary matrices representing linear maps $G_i : \mathbb{C}^2 \longrightarrow \mathbb{C}^2$.
Since the qubit state is the same in either the computational or the tensor basis (Eq.~\eqref{eq:basis_computational_1}), the action of the gates is also the same.
Using the line representation in Eq.~\eqref{eq:1_line}, the action of $G_1$ is a $2 \times 2$ matrix multiplication pictorially represented by
\begin{equation}
    \begin{tabular}{c}
    \includegraphics[height=2.3cm]{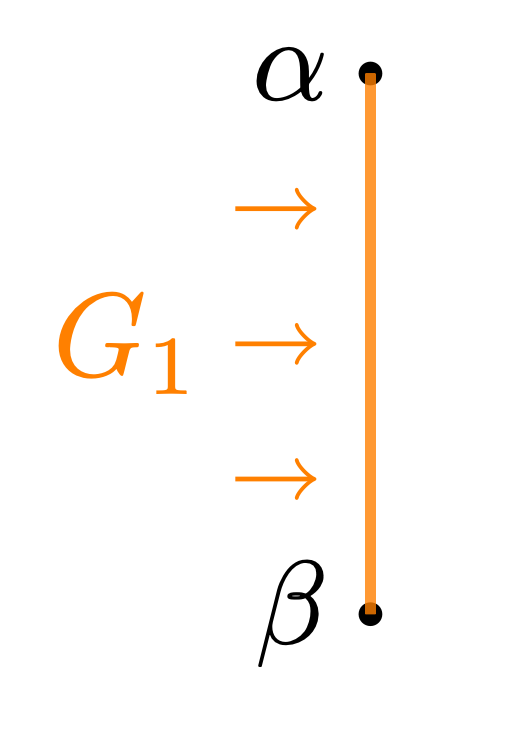}
    \end{tabular}
\end{equation}
In the case of a two-qubit system, local operations $G_{ij} \in \mathcal{SU}(4)$ are described in standard notation by a $4 \times 4$ unitary matrices, which can be written as a single Kronecker product of two single-qubit gates, namely $G_L = G_1 \boxtimes G_2$, as in Eq.~\eqref{eq:2qubits_gate}. 
This local gate acts, in standard notation, as a linear map $ G_L : \mathbb{C}^4 \longrightarrow \mathbb{C}^4$ and contains 16 potentially non-zero entries (cf. Eq.~\eqref{eq:2qubits_gate}).
However, the Property~\ref{prop:bilinear_tensor_linear} tells us that this linear map was equivalent to a bilinear map before we embedded the spaces. This means that the $G_L$ operation on tensors of order 2 can be defined as a bilinear map
\begin{equation}
        G_L = \left(G_1,G_2 \right): \mathbb{C}^2 \times \mathbb{C}^2 \longrightarrow \mathbb{C}^2 \times \mathbb{C}^2.
\end{equation}
Its action on the most general two-qubit state $\ket{\psi} = \ket{v_1}_1 \otimes \ket{v_2}_2 + \ket{w_1}_1 \otimes \ket{w_2}_2$ is
\begin{multline}
    G_L \ket{\psi} = (G_1, G_2) \cdot (v_1 \otimes v_2 + w_1 \otimes w_2)\\
    = G_1v_1 \otimes G_2v_2 + G_1w_1 \otimes G_2 w_2\\
    = G_1 (v_1 v_2^T + w_1 w_2^T) G_2^T\\
    = G_1 (v_1 \otimes v_2 + w_1 \otimes w_2) G_2^T,
    \label{eq:local_example_2}
\end{multline}
where we avoided the ket symbol. Using the cube representation in Eq.~\eqref{eq:2_cube}, the action of $G_L = (G_1, G_2)$ is
\begin{equation}
    \begin{tabular}{c}
    \includegraphics[width=6.5cm]{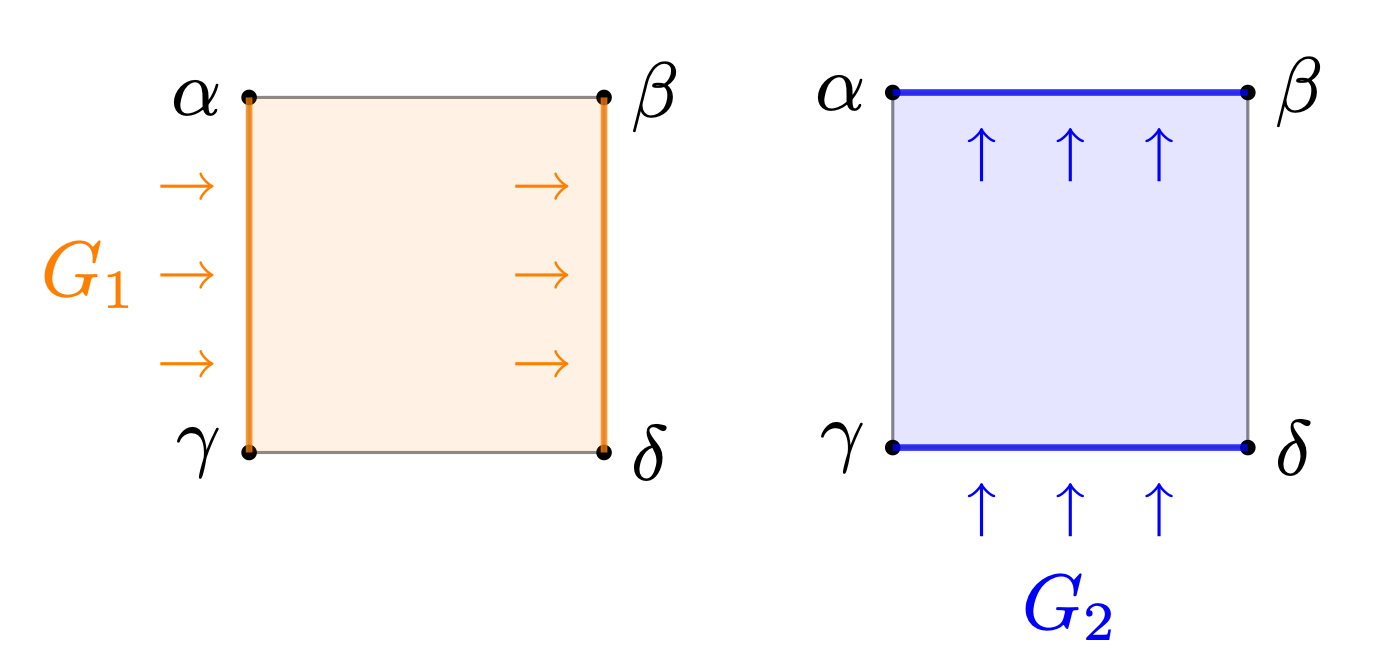}
    \end{tabular}
\end{equation}
where $G_i$ are applied sequentially. The two actions correspond to $ \left( G_1, \mathbb{1} \right)$ and $ \left( \mathbb{1}, G_2 \right)$ respectively and are obviously commutative.
$G_1$ acts horizontally on the vertical edges corresponding to the first qubit $\ket{q}_1$, and $G_2$ acts vertically on the horizontal edges corresponding to the second qubit $\ket{q}_2$.
The formalism can be extended to a three-qubit local gate, which in standard notation is an $8 \times 8$ unitary matrix derived from $G_L = G_1 \boxtimes G_2 \boxtimes G_3$. This contains 64 potentially non-zero complex entries, which is absolutely greater than the number of free parameters after considering unitary and separability constraints.
In terms of an operation on tensors, the same gate can be viewed as a single triplet expressing a multilinear map
\begin{equation}
        G_L = (G_1, G_2, G_3)
        : \left( \mathbb{C}^2\right)^{\times 3} \longrightarrow \left( \mathbb{C}^2\right)^{\times 3}
\end{equation}
acting on an order-3 tensor and preserving its rank. 
Instead of an $8 \times 8$ unitary matrix, we now have a triplet of $2 \times 2$ unitary matrices $G_i \in \mathcal{SU}(2)$, each acting in the dimension corresponding to the qubit $\ket{q}_i$, with a total of $12$ non-zero entries. 
Using the cube representation in Eq.~\eqref{eq:cube}, the sequentially application of $(G_1, \mathbb{1}, \mathbb{1})$, $(\mathbb{1}, G_2, \mathbb{1})$ and $(\mathbb{1}, \mathbb{1}, G_3)$ can be pictorially represented  as
\begin{equation}
\begin{tabular}{c}
    \includegraphics[width=7.5cm]{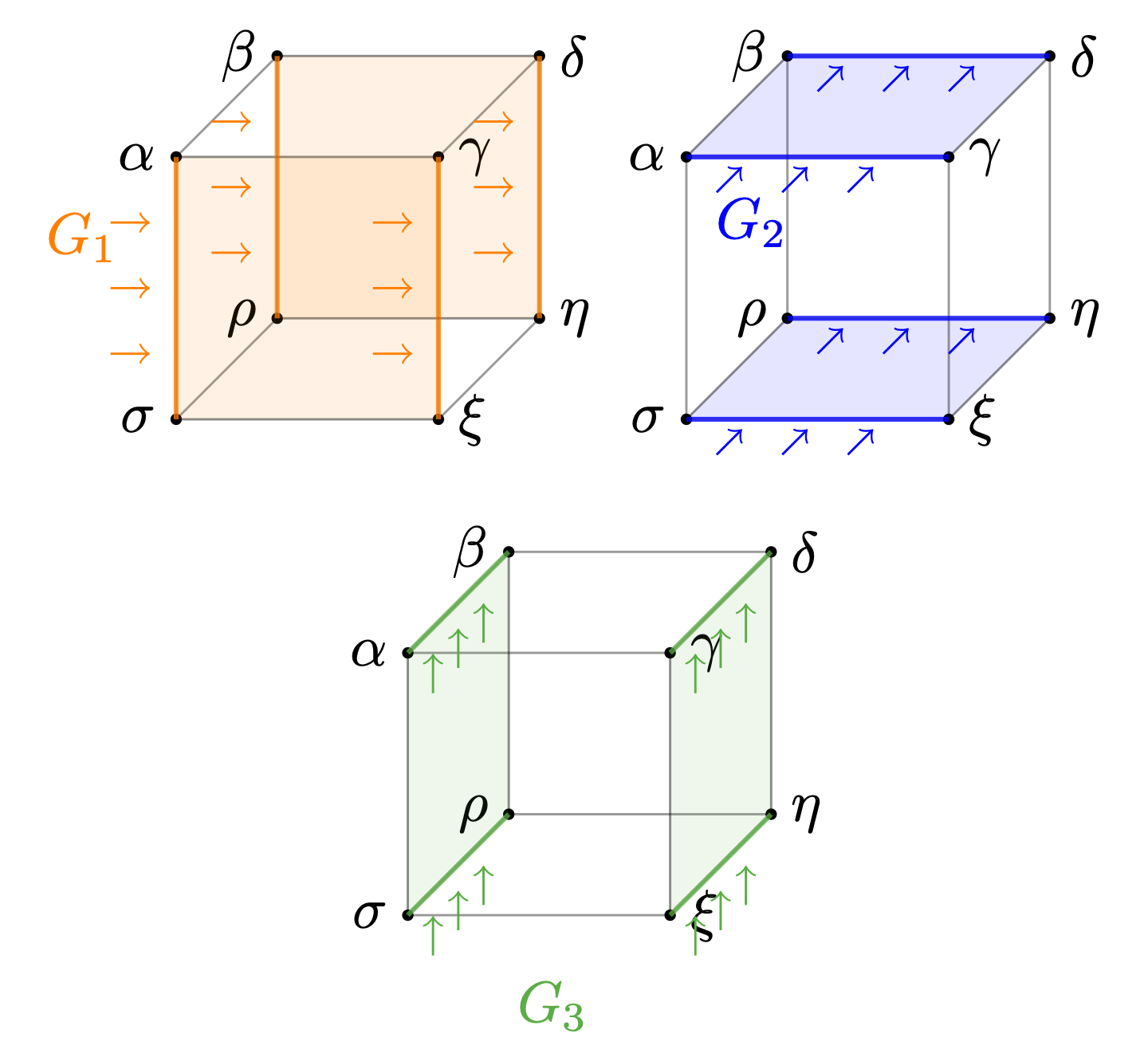}
\end{tabular}
\end{equation}
where $G_1$ acts on vertical edges oriented downward coherently with the dimension introduced by $\ket{q}_1$, $G_2$ acts  on horizontal edges oriented to the right as $\ket{q}_2$, and $G_3$ on horizontal edges oriented in depth as $\ket{q}_3$.
In summary, the introduction of an explicit tensorial formalism translates the gate application from an exponentially large matrix multiplication to a sequential application of small $2 \times 2$ matrices.

The multilinear definition of local multi-qubit gates has some advantages over the matrix representation.
(1) It preserves the separability property of the operation by keeping clear which gate acts on which space, and (2) it does not increase the number of non-trivial elements. In fact, by the usual convention, $G_L = G_1 \boxtimes \dots \boxtimes G_n$ is a $2^n \times 2^n$ unitary matrix with potentially $4^n$ non-zero entries, while $G_L = (G_1, \dots, G_n)$ is a multilinear map containing up to $4n$ non-zero entries.
Furthemore, the number of necessary parameters needed to identify the action of $G_L$ is generally not manifest in the standard notation, and the inverse step (from the $2^n \times 2^n$ matrix to the $n$-upla multilinear map) is not immediate at all.
In addition to minimising the number of non-trivial elements in the gate representation, (3) this explicitly tensorial formalism also minimises the number of matrix multiplications that has to be performed to compute the transformation of a state.
A local three-qubit gate requires $64$ multiplications and 56 sums in standard notation, but only 48 multiplications and 24 sums in tensor notation.
For a general value of $n$, standard notation requires $4^n$ multiplications and $2^n (2^n-1)$ sums and tensor notation requires $n 2^{n+1}$ multiplications and $n 2^{n}$ sums (for each qubit we have $2^{n-1}$ edges to act on, and each edge requires 4 multiplications and 2 sums).

\subsubsection{Controlled gates as quasi-multilinear maps}
We have seen that local operations can be described as multilinear maps acting on tensors instead of matrices acting on vectors.
However, a local operation transforms a quantum state while keeping the degree of entanglement constant. This means that $G_L$ cannot create or destroy entanglement between the two qubits. 
This also implies that the Schmidt number $s$ in Eq.~\eqref{eq:Schmidt_state} must remain the same.
Since the Property~\ref{prop:rank_multilinear} states that a multilinear map, composed by unitary linear maps, keeps the rank constant,
and in Eq.~\eqref{eq:Schmidt_state} $s$ is a rank, then the gate locality can be precisely interpreted as the Property~\ref{prop:rank_multilinear}. 
However, this means that non-local operations cannot be implemented by multilinear maps.

The general case of non-local gates is analysed in Section~\ref{Sec:non_local_gates}, and we consider here the case of controlled quantum gates, denoted as $\Lambda^n_{c}(G)$, where $n$ is the total number of qubits in the system, $c$ is the number of control qubits, and $G \in \mathcal{SU}(2)$ is the gate applied to the target qubit.
First we can note that controlled gates are by definition rank two operations, because we can always fix two ortonormal states of the control qubit register, which determine whether the gate $\Lambda^n_c(G)$ acts on the target qubit with $G$ or with $\mathbb{1}$.
Let us first consider a fully controlled gate $\Lambda^{n}_{n-1}(G)$ and define $\ket{\phi_c} = \ket{c_1}_1 \otimes \dots \otimes \ket{c_{n-1}}_{n-1}$, where $c_i \in \{0,1\}$, the state of the control qubits that satisfies the control condition, and $\ket{\phi_c}^{\perp}$ the orthonormal one.
Then, in Dirac notation, the gate can be written as
\begin{multline}
    \Lambda^n_{n-1}(G) = \op{\phi_c}{\phi_c} \otimes G + \ket{\phi_c}^{\perp}\bra{\phi_c}^{\perp} \otimes \mathbb{1}\\
    := M_c \otimes G + M_c^{\perp} \otimes \mathbb{1},
    \label{eq:full_controlled_gate}
\end{multline}
where we define the operator $M_c^{(\perp)}$ which projects the state along $\ket{\phi_c}$ ($\ket{\phi_c}^{\perp})$.
The $2^n \times 2^n$ representation matrix in standard notation can be found replacing $\otimes$ with $\boxtimes$.
The non-fully controlled case can be easily obtained by replacing the identity gate $\mathbb{1}$, in the tensor product, to the correct position corresponding to the qubit on which the gate does not act.
The linear map in Eq.~\eqref{eq:full_controlled_gate} can be also described by a sum of two multilinear maps
\begin{equation}
    \Lambda^n_{n-1}(G) = \left( M_c, G \right) + \left( M_c^{\perp}, \mathbb{1} \right),
    \label{eq:full_controlled_quasi}
\end{equation}
using again the Property~\ref{prop:bilinear_tensor_linear} for both terms.
Note that the sum of multilinear maps is not a multilinear map of sums, and this is reflected in the fact that a non-local gate, as a fully controlled gate, cannot correspond to a multilinear map, because it has to be able to destroy or generate entanglement.
In the case of $n=2$ qubits, the action of a controlled gate on a tensor state is
\begin{equation}
    \Lambda^2_1(G) \ket{\psi}  = M_c \psi G^T + M_c^{\perp} \psi  \mathbb{1}^T, 
    \label{eq:control_map_2}
\end{equation}
in a similar manner of Eq.~\eqref{eq:local_example_2}.

However, we can interpret the controlled-gate action in a more intuitive way by noting that there is only a multilinear map which actually change the state of the tensor, while the other is trivial and necessary only to guarantee the unitarity.
In practice, instead of computing the two multilinear map applications and finally summing the results (as in Eq.~\eqref{eq:control_map_2} for $n=2$), we can compute the action by performing the matrix multiplication only on the dimension of the hypercube corresponding to the target qubit, and only on the part of the tensor that satisfies the control condition.
We define this kind of action a \textit{quasi-multilinear map}.

\begin{defn}[Quasi-multilinear map]
    A quasi-multilinear map is an $n$-upla
    \begin{equation}
    \Lambda^n_{n-1}(G) := \left(C_{q_1}, \dots , C_{q_{n-1}} , G \right),
    \label{eq:quasi_multilinear_def}
    \end{equation}
    which represents a controlled quantum gate. It contains the set of conditions $C_{q_i}$, which determines the half of the tensor satisfying $\ket{q}_i = \ket{c_i}_i$, and the gate $G\in \mathcal{SU}(2)$, which acts on the target qubit but only on the part of the tensor that satisfies all the conditions. In the case of a fully controlled gate, the condition $C_{q_k}$ is replaced by $\mathbb{1}$ for each $\ket{q}_k$ that is not a control qubit.
\end{defn}

To explain as a quasi-multilinear map act on a tensor, consider a two-qubit controlled gate $\Lambda^2_1(G)$ which admits 4 possible conditions corresponding to the quantum circuits
\begin{equation}
	\begin{tabular}{c}
	\Qcircuit @C=0.6em @R=0.6em{
	& \ctrlo{1} & \qw &  &   \ctrl{1} & \qw & & \gate{G} & \qw & & \gate{G} & \qw\\
	&  \gate{G} & \qw &  &  \gate{G}& \qw & & \ctrlo{-1} & \qw & & \ctrl{-1} & \qw\\
	} \end{tabular}
	\label{eq:circ_CG}
\end{equation}
depending on the control state $\ket{\phi_c} \in \{\ket{q}_1 = \ket{0}_1, \ket{q}_1 = \ket{1}_1, \ket{q}_2 = \ket{0}_2, \ket{q}_2 = \ket{1}_2\}$.
So, using the quasi-multilinear map definition in Eq.~\eqref{eq:quasi_multilinear_def}, we can act on the tensor $\ket{\psi}$ by first selecting the edge that satisfies the control condition $C_{q_1}$ or $C_{q_2}$. The four possibilities can be pictorially represented by 
\begin{equation}
\begin{tabular}{c}
	\includegraphics[width=7.4cm]{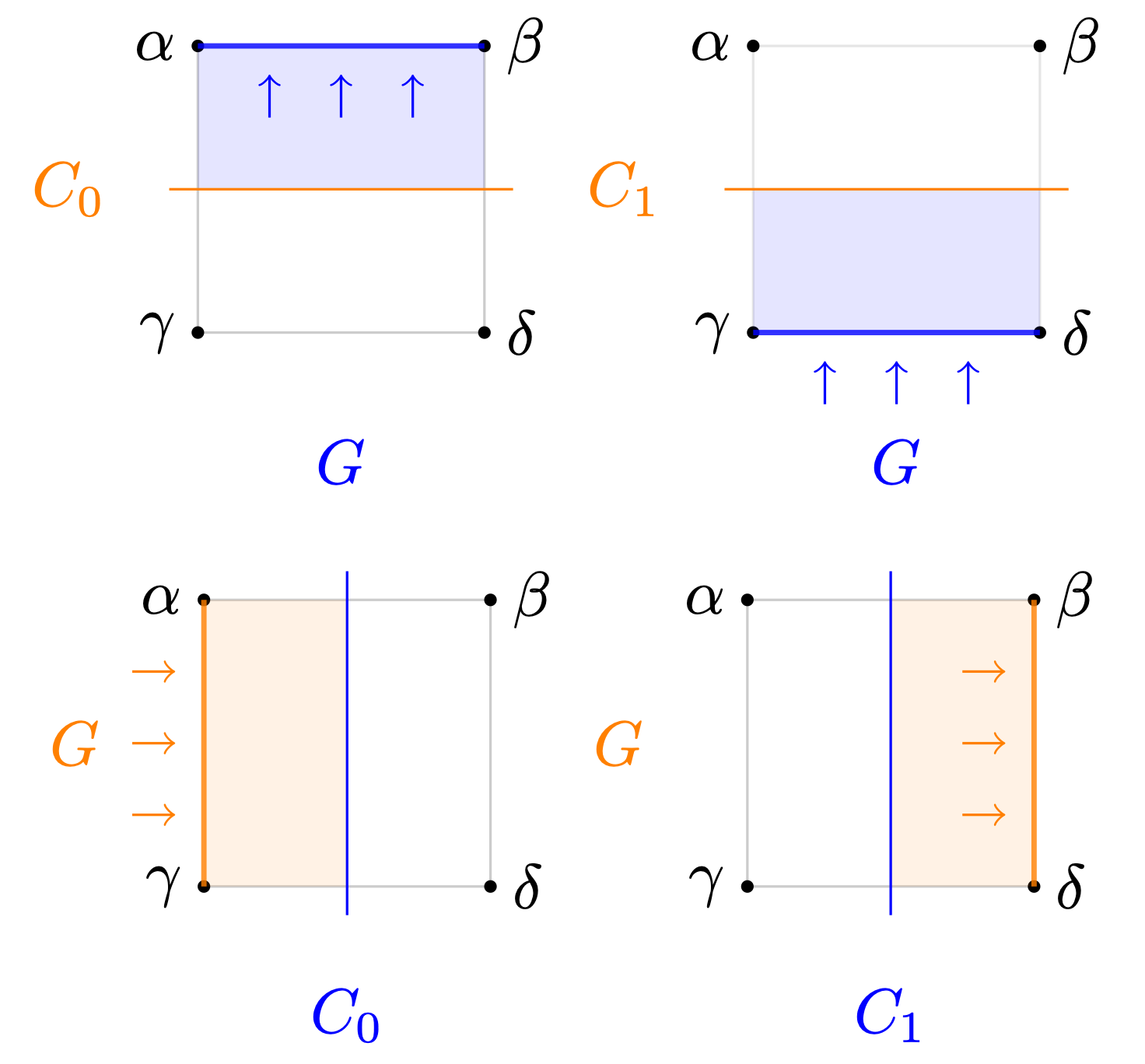}
\end{tabular}
\label{eq:cube_CG}
\end{equation}
and in terms of quasi-multilinear maps they correspond to $\Lambda^2_1(G) \in \left\{ (C_0,G), (C_1,G), (G,C_0), (G, C_1) \right\}$ respectively.
For example, the second circuit in Eq.~\eqref{eq:circ_CG}, for $G = X$, is the CNOT gate which corresponds to either the sum of bilinear maps $\text{CNOT} = \left( M_0, \mathbb{1} \right) +  \left( M_1,X \right)$, acting as
\begin{multline}
    \text{CNOT} \ket{\psi} = M_0 \begin{pmatrix}
    \alpha & \beta \\ \gamma & \delta
    \end{pmatrix} \mathbb{1}^T + M_1\begin{pmatrix}
    \alpha & \beta \\ \gamma & \delta
    \end{pmatrix} X^T \\
    = \begin{pmatrix}
    \alpha & \beta \\ 0 & 0
    \end{pmatrix} + \begin{pmatrix}
    0 & 0 \\ \delta & \gamma
    \end{pmatrix} = \begin{pmatrix}
     \alpha & \beta \\ \delta & \gamma
    \end{pmatrix},
\end{multline}
or to the quasi-multilinear $\text{CNOT} = (C_1, X)$ which inverts the vertices $\gamma$ and $\delta$ in the upper right cube of Eq.~\eqref{eq:cube_CG}.

In the case of three qubits the approach is the same and the control condition fixes the part of the cube on which we have to apply $G$. If we have a single control qubit, the control condition fixes a face. For example, the gate $\Lambda^3_1(X) = \text{CNOT} \otimes \mathbb{1}$ acts with the $X$ gate on the second qubit only if the first satisfies $\ket{q}_2 = \ket{1}_2$.
This corresponds to the sum of two trilinear maps
\begin{equation}
    \text{CNOT} \otimes \mathbb{1} = \left(M_0 , \mathbb{1}, \mathbb{1} \right) + \left(M_1, X, \mathbb{1}\right)
\end{equation}
or to the single quasi-multilinear map
\begin{equation}
    \text{CNOT} \otimes \mathbb{1} = \left(C_1, X, \mathbb{1} \right).
    \label{eq:CNOT_I}
\end{equation}
In Eq. \eqref{eq:CNOT_I}, the condition $C_1$ fixes the half of the tensor, corresponding to $\ket{q}_1 = \ket{1}_1$, on which we apply $X$. Thus, we do the matrix multiplication only on the bottom face of the cube
\begin{equation}
	\begin{tabular}{c}
	\includegraphics[width=7.5cm]{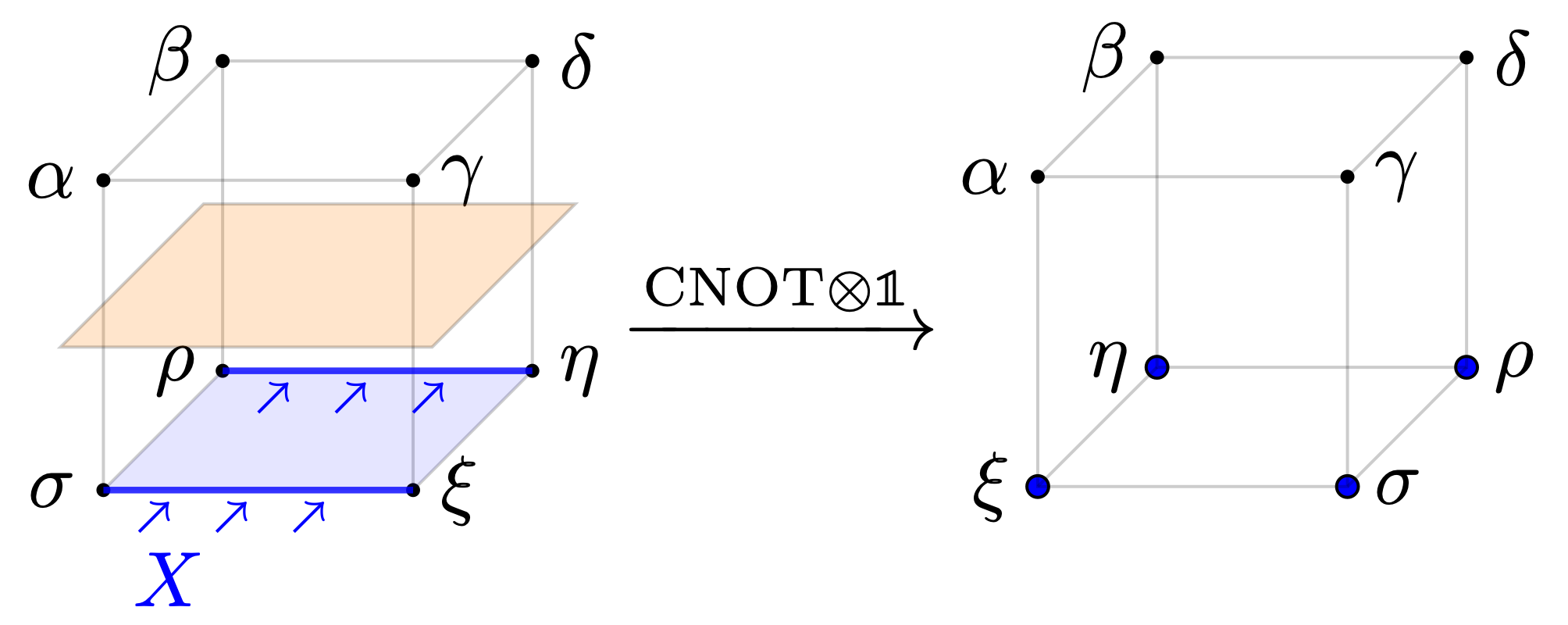}
	\end{tabular}
\end{equation}
where the orange face halves the tensor due to the $C_1$ condition. 
The key point is that each control condition halves the dimension of the tensor that evolves non-trivially. 
Consider the Toffoli gate $\Lambda^3_{2}(X) = \text{CCNOT}$, acting on the first qubit, as another example.
It acts on $\ket{q}_1$ if the other two satisfy $\ket{q}_2\otimes \ket{q}_3 = \ket{1}_2\otimes \ket{1}_3$ and can be expressed as the sum of the two multilinear maps
\begin{equation}
    \Lambda^3_2(X) = (X,M_{c}) +  (\mathbb{1},M_c^{\perp}),
\end{equation}
where $M_c = \op{11}{11}$ and $M_c^{\perp} = \op{00}{00} + \op{01}{01} + \op{10}{10}$, or as the quasi-multilinear map
\begin{equation}
    \Lambda^3_2(X) = \left(X,C_1, C_1\right),
\end{equation}
which acts on the order-3 tensor by changing only the part of the tensor which satisfies both conditions.
We pictorially represent it as
\begin{equation}
	\begin{tabular}{c}
	\includegraphics[width=7.5cm]{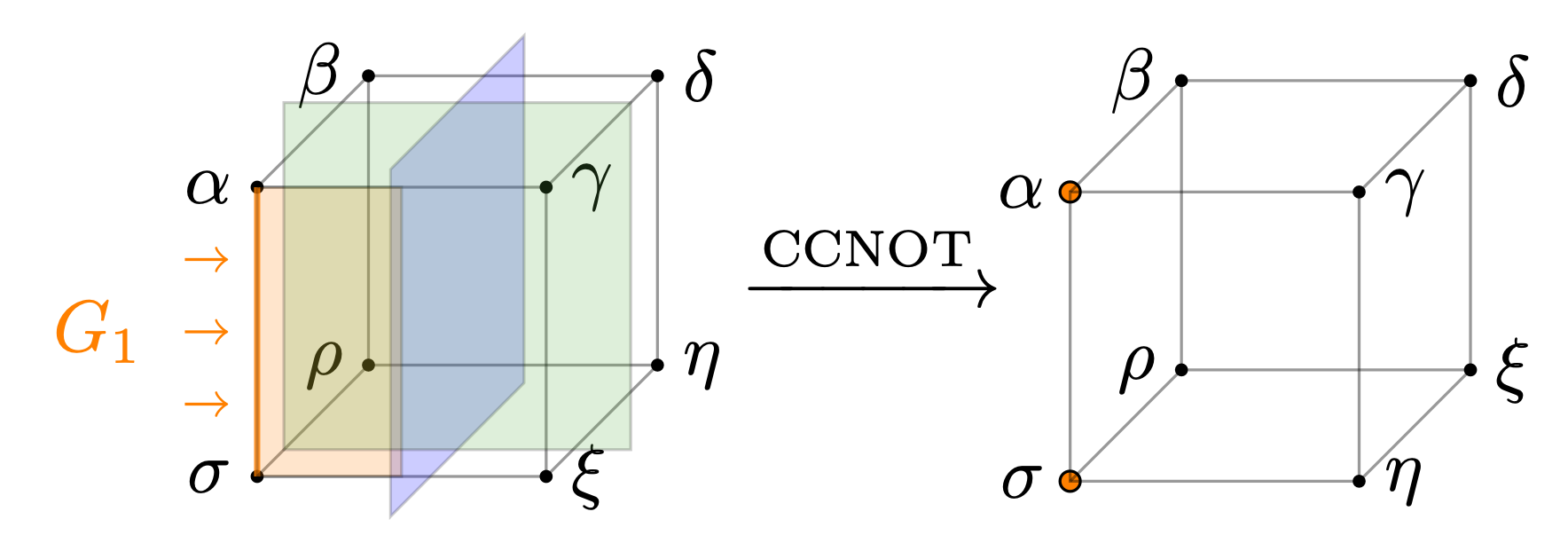}
	\end{tabular}
\end{equation}
where the blue and the green surfaces fix the first (left face $\ket{q}_2 = \ket{1}_2$) and the second (front face $\ket{q}_3 = \ket{1}_3$) conditions respectively, and the $X$ gate is applied on the first qubit but only on the edge fixed by the intersection of the two selected faces.
This description can not be easily extended to the canonical vector notation.

The point is that more are the control conditions, and fever is the dimension of the tensor on which we have to act with the single-qubit gate $G$.
For a non-fully controlled operation, each control divides the tensor into two parts, corresponding to the satisfied and the unsatisfied condition. For example, for three qubits $\Lambda^3_1(G)$ acts with $G$ on one face and $\Lambda^3_2(G)$ on one edge.
So for a fully controlled gate $\Lambda^n_{n-1}$ the $n-1(G)$ conditions fix the single edge we have to act on, so we only need 4 multiplications and 2 sums.
To see the explicit tensor notation in action, we have reported in Appendix~\ref{sec:algorithm} the state evolution of the famous quantum teleportation algorithm.
 
\subsubsection{Non-local gate}
\label{Sec:non_local_gates}
In the case of general non-local operations, there is no interpretation equivalent to that proposed for controlled gates.
So they cannot be expressed as quasi-multilinear maps.
However, we can express a non-local $n$ qubit gate using the expression in Eq.~\eqref{Eq:nqubit_gate_matrix}, which can be interpreted as a sum of $r$ multilinear maps
\begin{equation}
    G = \sum_{i=1}^r \left( G_1^i , \dots , G_n^i \right)
    \label{Eq:nqubit_gate_multilinear}
\end{equation}
again using the Property~\ref{prop:bilinear_tensor_linear} for each term.
The explicit tensor notation thus transforms the $2^n \times 2^n$ matrix multiplication into a set of $n$ small $2 \times 2$ matrix multiplications, and finally sums the results.
This reduces the number of multiplications needed. In fact, in standard notation we have to compute $4^n$ multiplications, whereas to do the same transformation in tensor notation we need $r n 2^{n+1}$ multiplications.
Note that the rank is bounded by $r \leq 2^n$, and for any polinomial rank $r = \mathcal{O}(\text{poly}(n))$ the tensor approach guarantees a complexity speedup, since $\mathcal{O}( \text{poly} (n) 2^{n+1}) \ll \mathcal{O}(4^n)$.

\section{Conclusions}
Principles from tensor theory and multilinear maps are integrated into the description of quantum computation by the introduced tensorial formalism.
This provides a framework that coherently describes the qubit states and the operations on them, and could improve our intuitive understanding of their behaviour.
This is suggested in particular by the preservation of the structure of Hilbert space, which provides a link between mathematical foundations and quantum phenomena in computation.
The proposed formalism could represent a step forward in the field of quantum computation by suggesting some practical advantages.
One of these is the possible reduction in the computational complexity required to predict the effect of quantum gates on multi-qubit states via classical computation. 
This could be exploited to predict the evolution of qubit states along the quantum circuit, as done for quantum teleportation in the Appendix \ref{sec:algorithm}, but involving a more substantial number of qubits.
Moreover, this establishes a link between the quantum gate's ability to generate entanglement and its mathematical description in terms of multilinear maps acting on tensors. The ability to generate entanglement is linked to the ability to change the rank of the tensor representing the qubit system, and this in turn is linked to the rank of the gate representation, which does not provide any useful information in standard matrix notation.
Another possible future investigation is to look for a connection between the rank of the quantum gate and the circuit-based complexity of the operation.
Finally, exploring and quantifying entanglement is a central challenge in quantum information theory, and our framework provides, to our knowledge, a new lens through which to address this challenge. 
The entanglement between different qubits, and more generally that of composite quantum systems, is correlated with the direction of the cross-connections between the spaces connected by the tensor product, suggesting a possible new perspective on the measurement of entanglement.




\vspace{1cm}
\textit{Acknowledgements} \----
I would like to thank Professor of Mathematics Alessandra Bernardi for insightful discussions.
This work was supported by the Department of Physics of the University of Trento (\url{https://www.physics.unitn.it/}) and the Trento Institute for Fundamental Physics and Applications (\url{https://www.tifpa.infn.it/}).


%

\input{Appendice}

\end{document}

%% file: 1_line.tex
    \begin{tikzpicture}[scale=0.6]
        \coordinate (B) at (-1,-1,1);
        \coordinate (C) at (-1,1,1);
        \draw[fill=black] (C) circle (1pt) node[anchor=east] {$\ket{0}$};
        \draw[fill=black] (B) circle (1pt) node[anchor=east] {$\ket{1}$};
        \draw[opacity=0.4] (B)--(C)--cycle;
    \end{tikzpicture}

%% file: 2_square.tex
    \begin{tikzpicture}[scale=0.8]
        \coordinate (A) at (1,-1,1);
        \coordinate (B) at (-1,-1,1);
        \coordinate (C) at (-1,1,1);
        \coordinate (D) at (1,1,1);
        \draw[fill=black] (C) circle (1pt) node[anchor=east] {$\ket{00}$};
        \draw[fill=black] (D) circle (1pt) node[anchor=west] {$\ket{01}$};
        \draw[fill=black] (B) circle (1pt) node[anchor=east] {$\ket{10}$};
        \draw[fill=black] (A) circle (1pt) node[anchor=west] {$\ket{11}$};
        \draw[opacity=0.4] (A)--(B)--(C)--(D)--cycle;
    \end{tikzpicture}

%% file: 3_cube.tex
    \begin{tikzpicture}[scale=0.8]
        \coordinate (A) at (1,-1,1);
        \coordinate (B) at (-1,-1,1);
        \coordinate (C) at (-1,1,1);
        \coordinate (D) at (1,1,1);
        \coordinate (E) at (1,1,-1);
        \coordinate (F) at (-1,1,-1);
        \coordinate (G) at (-1,-1,-1);
        \coordinate (H) at (1,-1,-1);

        \draw[fill=black] (C) circle (1pt) node[anchor=east] {$\ket{000}$};
        \draw[fill=black] (F) circle (1pt) node[anchor=east] {$\ket{001}$};
        \draw[fill=black] (D) circle (1pt) node[anchor=west] {$\ket{010}$};
        \draw[fill=black] (E) circle (1pt) node[anchor=west] {$\ket{011}$};
        \draw[fill=black] (B) circle (1pt) node[anchor=east] {$\ket{100}$};
        \draw[fill=black] (G) circle (1pt) node[anchor=east] {$\ket{101}$};
        \draw[fill=black] (A) circle (1pt) node[anchor=west] {$\ket{110}$};
        \draw[fill=black] (H) circle (1pt) node[anchor=west] {$\ket{111}$};

        \draw[opacity=0.4] (A)--(B)--(C)--(D)--cycle;
        \draw[opacity=0.4] (E)--(F)--(G)--(H)--cycle;
        \draw[opacity=0.4] (A)--(H);
        \draw[opacity=0.4] (B)--(G);
        \draw[opacity=0.4] (D)--(E);
        \draw[opacity=0.4] (C)--(F);
    \end{tikzpicture}

%% file: Appendice.tex
\section{The quantum teleportation algorithm}
\label{sec:algorithm}
The quantum teleportation algorithm, first formulated by Bennett et al. in 1993~\cite{bennett1993teleporting}, exemplifies the profound departure from classical notions of information transfer, involving the delicate interplay of entanglement, superposition and measurement, which give quantum particles the capacity to exist in multiple states simultaneously and to establish non-local correlations.
Quantum teleportation allows to transmit an unknown quantum state $\ket{\phi}$ from a sender (Alice) to a receiver (Bob) who share an intermediary entangled pair of particles at the beginning and can use only local operations and classical communication.
The quantum algorithm is represented by the following quantum circuit
\begin{equation}
\begin{tabular}{c}
\Qcircuit @C=1em @R=1em{
	\lstick{ \ket{\phi}_1} & \barrier{2} , \qw & \qw \barrier{2} &  \ctrl{1} \barrier[-1.3em]{2}  & \gate{H} \barrier[-1.4em]{2}  & \meter{} &  \control \cw \cwx[2]\\
	\lstick{ \ket{0}_2} & \gate{H} & \ctrl{1} & \targ{} & \qw & \meter{} & \cw & \control \cw \cwx[1]\\
 	\lstick{ \ket{0}_3} & \qw & \targ{} & \qw & \qw & \qw & \gate{Z} & \gate{X} &  \rstick{\ket{\phi}} \qw
	}
\end{tabular}
\label{circ:teleportation}
\end{equation}
consisting of (1) the Bell state preparation, (2) the local operation performed by Alice (quantum register $\ket{q}_1 \otimes \ket{q}_2$), and finally (3) the classical communication and the local operation performed by Bob (quantum register $\ket{q}_3$).
Note that the circuit contains the quantum gates $Z = \sigma_z$, $X = \sigma_x$ and the Hadamard $H$.
In the following sections, we prove the power of the three-qubit explicit tensor notation in the study of state evolution and measurement interpretation on the quantum state evolving along the circuit~\eqref{circ:teleportation}.
\begin{figure*}
\begin{tabular}{c}
	\includegraphics[width=17.9cm]{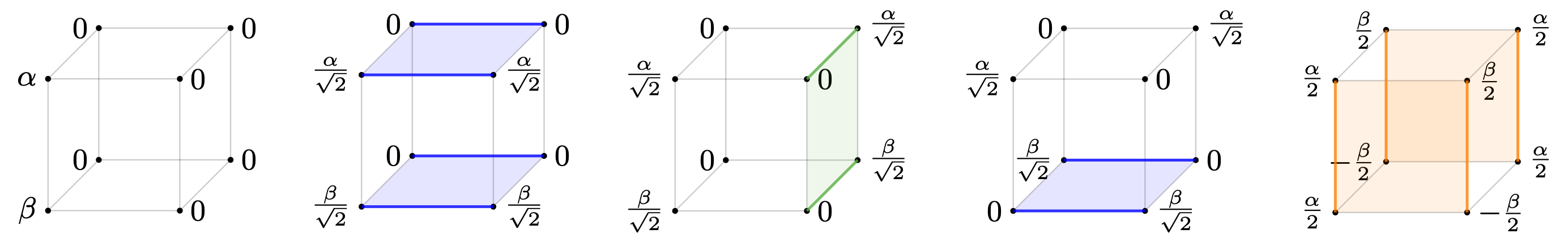}
\end{tabular}
\caption{Cube representation of the three-qubit state, expressed as an order-3 tensor, along the quantum circuit~\eqref{circ:teleportation}. From left to right, the cubes represent $\psi^{(0)}$, $\psi^{(1)}$, $\psi^{(2)}$, $\psi^{(3)}$ and $\psi^{(0)}$, where the coloured faces correspond to the action of the multilinear or quasi-multilinear maps generated by the quantum operations along the circuit.}
\label{fig:state_evolution}
\end{figure*}

\subsection{Quantum state evolution}
The state evolution along the quantum circuit~\eqref{circ:teleportation} is described below in Dirac notation and represented in Fig.~\ref{fig:state_evolution} in tensor notation, but we omit the vector representation, which would not be intuitive and would involve $8 \times 8$ matrix multiplications, thus requiring more computational effort.
The initial state is 
\begin{equation}
	\ket{\psi}^{(0)} = \ket{\phi} \otimes \ket{0} \otimes \ket{0} = \left( \alpha \ket{0} + \beta \ket{1} \right) \otimes \ket{0} \otimes \ket{0}
	\label{eq:psi_0}
\end{equation}
which consists of an order-3 tensor and can be represented as the first cube in Fig.~\ref{fig:state_evolution}.
This has non-zero entries in the first qubit dimension (the vertical one) and in the edge coming from the intersection between the face $\ket{q}_2 = \ket{0}_2$ and the face $\ket{q}_3 = \ket{0}_3$.
This state has rank
\begin{equation}
	\rank\left(\psi^{(0)}\right) = 1,
\end{equation}
which is emphasised by both Dirac and tensor notation, and therefore corresponds to a separable state.

The first operation in~\eqref{circ:teleportation} is a local gate which acts with a Hadamard on the second qubit, i.e. $G^{(1)} = \mathbb{1} \otimes H \otimes \mathbb{1}$ and transforms the state in Eq. \eqref{eq:psi_0} into
\begin{equation}
	\ket{\psi}^{(1)} = \ket{\phi} \otimes \ket{+} \otimes \ket{0}.
	\label{eq:psi_1}
\end{equation}
Using the tensor formalism, the operation is the single trilinear map $G^{(1)} := (\mathbb{1}, H, \mathbb{1})$ which acts non-trivially on the horizontal edges corresponding to $\ket{q}_2$ and transforms the cube into the second one shown in Fig.~\ref{fig:state_evolution}.
The operation $G^{(1)}$ is local, so the tensor preserves the rank, namely
\begin{equation}
	\rank \left( \psi^{(1)} \right) = 1.
\end{equation}

The second operation in~\eqref{circ:teleportation} is a CNOT that acts on the third qubit if the second is in $\ket{q}_2 = \ket{1}_2$, and thus corresponds to $G^{(2)} = \mathbb{1} \otimes \text{CNOT}$ and maps the Eq.~\eqref{eq:psi_1} to
\begin{equation}
	\ket{\psi}^{(2)} = \ket{\phi} \otimes \frac{\ket{0} \otimes \ket{0} + \ket{1} \otimes \ket{1}}{\sqrt{2}} = \ket{\phi} \otimes \ket{\Phi^+}.
	\label{eq:psi_2}
\end{equation}
In the above equation we have defined the Bell state $\ket{\Phi^+}$, which is the result of the entangled state preparation required for the quantum teleportation protocol.
This second operation can be expressed as the quasi-trilinear map $G^{(2)} = (\mathbb{1}, C_1, X)$ which operates on a face giving the tensor represented by the third cube in Fig.~\ref{fig:state_evolution}.
The CNOT is a non-local gate that can change the rank, and in this case it has transformed the rank-1 tensor into a rank-2 tensor
\begin{equation}
	\rank \left( \psi^{(2)} \right) = 2.
\end{equation}

The preparation block is complete and Alice now transforms her qubits by first applying a CNOT which acts on the second qubit if $\ket{q}_1 = \ket{1}_1$. The corresponding operation is $G^{(3)} = \text{CNOT} \otimes \mathbb{1}$ and produces 
\begin{multline}
	\ket{\psi}^{(3)} = \alpha \ \ket{0} \otimes \frac{\ket{0} \otimes \ket{0} + \ket{1} \otimes \ket{1}}{\sqrt{2}} + \\
	+ \beta\ \ket{1} \otimes \frac{\ket{1} \otimes \ket{0} + \ket{0} \otimes \ket{1}}{\sqrt{2}}
	\label{eq:psi_3}
\end{multline}
from the state in Eq.~\eqref{eq:psi_2}.
The operation is again a controlled gate, which can be expressed as the quasi-multilinear map $G^{(3)} = (C_1, X, \mathbb{1})$ acting on the horizontal edges of the face $\ket{q}_1 = \ket{1}_1$. In terms of a hypermatrix, the operation and the resulting state are represented by the fourth cube in Fig.~\ref{fig:state_evolution}.
The gate is non-local and gives 
\begin{equation}
	\rank \left( \psi^{(3)} \right) = 4,
\end{equation}
thus increasing the degree of entanglement.

Finally, the last operation performed by Alice is $G^{(4)} = H \otimes \mathbb{1} \otimes \mathbb{1}$ and transformed Eq.~\eqref{eq:psi_3} into
\begin{multline}
	\ket{\psi}^{(4)}= \alpha  \frac{  \ket{0} \otimes\ket{+} \otimes \ket{0} +  \ket{0}\otimes\ket{-}\otimes\ket{1} }{\sqrt{2}} +\\
	+ \beta \frac{\ket{1}\otimes\ket{-}\otimes\ket{0} + \ket{1}\otimes\ket{+}\otimes\ket{1} }{\sqrt{2}}.
	\label{eq:psi_4}
\end{multline}
The quantum operation is local and corresponds to the multilinear map $G^{(4)} = (H, \mathbb{1}, \mathbb{1})$ acting on a tensor of order 3. This acts on the vertical edges corresponding to $\ket{q}_1$ and produces the tensor represented by the last cube in Fig.~\ref{fig:state_evolution}. This is local and therefore preserves the rank, so at the end of the quantum operations of the circuit~\eqref{circ:teleportation} we have a fully entangled state, i.e.
\begin{equation}
	\rank \left( \psi^{(4)} \right) = 4.
\end{equation}

\subsection{Measurement procedure and outcomes}
\begin{figure*}
\begin{tabular}{c}
	\includegraphics[width=17.9cm]{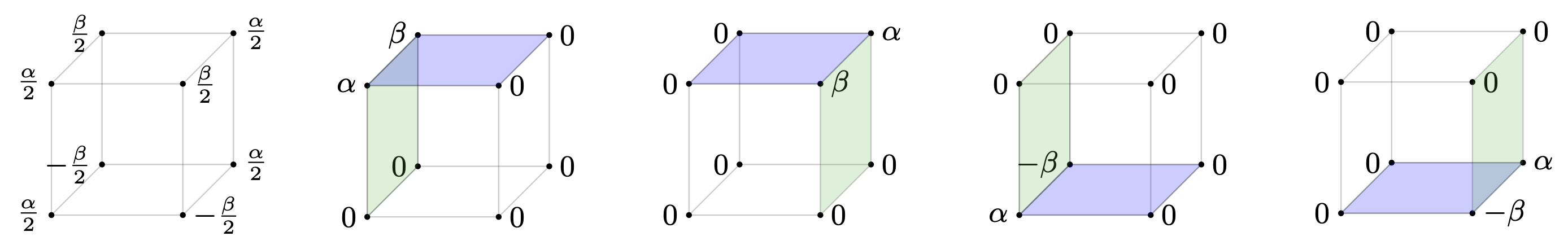}
\end{tabular}
\caption{Cube representation of the possible outcomes of the measurement procedure at the end of the quantum circuit~\eqref{circ:teleportation}. The first cube represents the state before the measurements. Then, from left to right, the cubes represent the results after Alice has measured her two qubits. The blue and green faces represent the result of measuring $\ket{q}_1$ and $\ket{q}_2$ respectively.  The four possibilities are given by the results $\ket{q}_1 \otimes \ket{q}_2 = \ket{0} \otimes \ket{0}$, $\ket{q}_1 \otimes \ket{q}_2 = \ket{0} \otimes \ket{1}$, $\ket{q}_1 \otimes \ket{q}_2 = \ket{1} \otimes \ket{0}$ and $\ket{q}_1 \otimes \ket{q}_2 = \ket{1} \otimes \ket{1}$ respectively. }
\label{fig:state_measure}
\end{figure*}
As mentioned in Sec.~\ref{sec:measurement}, the measurement of each qubit halves the dimension on the tensor by selecting the part of it which satisfies the obtain result. 
For this reason, measuring two out of three qubits collapses the state to an edge corresponding to the intersection of the two obtained results which correspond to an order-2 tensor, i.e. a matrix represented on a face of the cube.
The measurement procedure at the end of the circuit~\eqref{circ:teleportation} and the 4 possible obtains outcomes are represented in Fig.~\ref{fig:state_measure} where coloured faces are connected to the outcome measure of one qubit.
The first cube represents the quantum state obtained at the end of the quantum circuit, just before measure.
If we measure the first qubit in $\ket{q}_1 = \ket{0}_1$ we collapse the system in the upper face and with $\ket{q}_2 = \ket{0}_2$ we collapse the system in the left face. The two conditions together identify the upper edge on the left. This edge identifies the state for the third qubit, which in this case is $\ket{q}_3 = \ket{\phi}_3 = \alpha \ket{0}_3 + \beta \ket{1}_3$. If we measure $\ket{q}_1 = \ket{0}_1$ and $\ket{q}_2 = \ket{1}_2$ we identify the upper edge on the right, which means $\ket{q}_3 = \beta \ket{0}_3 + \alpha \ket{1}_3$, so Bob has to apply $X$ to recover $\ket{\phi}_3$. With $\ket{q}_1 = \ket{1}_1$ we identify the lower face and with $\ket{q}_2 = \ket{0}_2$ the left one, so the state is $\ket{q}_3 = \alpha \ket{0}_3 - \beta \ket{1}_3$ and Bob apply the $Z$ gate to recover $\ket{\phi}_3$. Finally measureing $\ket{q}_1 = \ket{1}_1$ and $ \ket{q}_2 = \ket{1}_2$ the result is at the bottom right edge and the final state is $\ket{q}_3 = - \beta \ket{0}_3 + \alpha \ket{1}_3$. By using the $XZ$ gate Bob gets $\ket{\phi}_3$. The post-measurement states in the four cases are shown in Fig. \ref{fig:state_measure}.